\documentclass[preprint,3p,twocolumn,authoryear]{elsarticle}

\usepackage[version=3]{mhchem} 
\usepackage{units}
\usepackage{siunitx}
\usepackage{amssymb}

\journal{}

\usepackage{etoolbox}

\makeatletter 
\def\ps@pprintTitle{  
\let\@oddhead\@empty 
\let\@evenhead\@empty  
\def\@oddfoot{\hfill\thepage}  
\def\@evenfoot{\thepage\hfill}} 
\makeatother

\usepackage{color}
\usepackage{hyperref, soul}
\definecolor{linkColor}{rgb}{1,0,0}
\definecolor{citeColor}{rgb}{1,0,0}
\hypersetup{pdfborder={0 0 0}, colorlinks=true,urlcolor=linkColor,citecolor=citeColor}

\renewcommand*{\today}{March 10, 2021}

\begin{document}

\begin{frontmatter}

\title{Fast Grain Mapping with Sub-Nanometer Resolution Using 4D-STEM \\with Grain Classification by Principal Component Analysis \\and Non-Negative Matrix Factorization}

\author[label1,label2]{Frances I. Allen\corref{cor1}}

\cortext[cor1]{Corresponding author}
\ead{francesallen@berkeley.edu} 

\address[label1]
{Department of Materials Science and Engineering, UC Berkeley, Berkeley, CA 94720, USA}
\address[label2]
{National Center for Electron Microscopy, Molecular Foundry, LBNL, Berkeley, CA 94720, USA}

\author[label1,label2]{Thomas C. Pekin\fnref{fn1}}
\fntext[fn1]{Current address: Institut für Physik \& IRIS Adlershof, Humboldt Universität zu Berlin, 12489 Berlin, Germany}

\author[label3]{Arun Persaud}
\address[label3]{Accelerator Technology and Applied Physics Division, LBNL, Berkeley, CA 94720, USA}

\author[label4]{Steven J. Rozeveld}
\author[label4]{\\Gregory F. Meyers}
\address[label4]
{Core R\&D - Analytical Sciences, The Dow Chemical Company, Midland, MI 48674, USA\vskip 10pt \textnormal{\today}}

\author[label2]{Jim Ciston}
\author[label2]{Colin Ophus}
\author[label1,label2]{Andrew M. Minor}

\begin{abstract}

High-throughput grain mapping with sub-nanometer spatial resolution is demonstrated using scanning nanobeam electron diffraction (also known as 4D scanning transmission electron microscopy, or 4D-STEM) combined with high-speed direct electron detection. An electron probe size down to \unit[0.5]{nm} in diameter is implemented and the sample investigated is a gold-palladium nanoparticle catalyst. Computational analysis of the 4D-STEM data sets is performed using a disk registration algorithm to identify the diffraction peaks followed by feature learning to map the individual grains. Two unsupervised feature learning techniques are compared: Principal component analysis (PCA) and non-negative matrix factorization (NNMF). The characteristics of the PCA versus NNMF output are compared and the potential of the 4D-STEM approach for statistical analysis of grain orientations at high spatial resolution is discussed.

\end{abstract}


\end{frontmatter}

\section{Introduction}

The last decade has borne witness to a surge in scanning transmission electron microscopy (STEM) experiments implementing scanning nanobeam electron diffraction (NBED), fueled by the development of high-speed, high-efficiency, direct electron detectors, as well as advanced computational methods. In scanning NBED the electron probe is rastered over 2D spatial co-ordinates ($x,y$) and an NBED pattern with 2D reciprocal space co-ordinates ($k_x,k_y$) is acquired at every dwell point. The result is a data set with co-ordinates in four dimensions, hence the alternative name for the technique, ``4D-STEM". 4D-STEM data sets contain a wealth of information that can be extracted by various means for a range of analytical purposes \citep{Ophus2019}. For example, virtual dark-field images can be reconstructed for any desired combination of diffraction peaks, enabling a multitude of virtual imaging experiments to be performed on the same data set post-acquisition \citep{Gammer2015}. By analyzing the spacing between diffraction peaks, measurements of crystal lattice strain can be made \citep{Beche2009}, and by acquiring 4D-STEM data during \textit{in situ} deformation experiments, strain evolution can be probed locally in time-discrete steps \citep{Pekin2018}. Converging the electron beam such that the Bragg diffraction disks significantly overlap enables electron ptychography experiments in which phase information is retrieved from the beam interference for resolution-enhanced imaging \citep{Jiang2018}. Electrostatic field mapping around individual atoms in 2D monolayers has also been demonstrated via analysis of the center of mass of the NBED intensity distributions \citep{Fang2019}. In the work presented here, 4D-STEM is implemented in combination with fast electron detection to explore new possibilities for high-throughput grain mapping at high spatial resolution.

Existing electron-beam-based grain mapping techniques are primarily electron backscatter diffraction (EBSD) in the scanning electron microscope (SEM) and scanning precession electron diffraction (SPED) in STEM. EBSD is the method of choice for grains in the micron size range and above, and by using a thin sample and adopting a transmission geometry to instead collect forward-scattered electrons (which emanate from a much smaller interaction volume), mapping of grains down to \unit[10]{nm} in size can be achieved \citep{Keller2012}. A key benefit of EBSD in the SEM is the ability to rapidly scan and index over fields of view up to thousands of square microns. However, the spatial resolution of SEM-based techniques is ultimately limited by the electron probe size. Grain mapping at high spatial resolution is therefore well-suited to STEM, where a sub-nanometer electron probe can be employed. In the precession-based SPED method (also a form of 4D-STEM), the STEM probe is rotated about a pivot point on the sample through a pre-defined tilt angle for each (x,y) co-ordinate in the scan. Hence SPED gives diffraction patterns averaged over a range of orientations enabling robust indexing, albeit requiring long scan times \citep{Rouviere2013,Midgley2015}. In contrast, in 4D-STEM without precession, a single diffraction pattern is acquired per dwell point, reducing the total scan time (and thus total electron dose) considerably, in addition to greatly simplifying the experimental setup. While 4D-STEM without precession cannot offer the same level of indexing precision as SPED (in terms of the enhanced interpretability of the intensities in the indexed spots due to the quasi-kinematical nature of diffraction patterns), it has for example been shown to be well-suited for grain orientation mapping of beam-sensitive materials \citep{Panova2016}. Moreover, as we show here, 4D-STEM without precessing the beam presents a powerful technique for grain orientation mapping when both high throughput and high spatial resolution are required. 4D-STEM using an aperture to form multiple beams is now also poised to enhance the technique even further \citep{Hong2020}. 

In the following, we apply 4D-STEM to an industrial catalyst comprising gold-palladium nanoparticles embedded in an silica support, using an aberration-corrected STEM equipped with a high-speed direct electron detector to map large sample sets with sub-nanometer resolution. 4D-STEM data sets are inherently large, especially when surveying large fields of view with a sub-nanometer scan spacing. Hence efficient methods for data processing with high levels of automation are essential. In our work, the diffraction peaks in the 4D-STEM data sets are first identified with sub-pixel precision using a disk registration algorithm developed previously \citep{Pekin2017}. Then, grain classification according to orientation is achieved \textit{a priori} by unsupervised feature learning, using principal component analysis (PCA) and non-negative matrix factorization (NNMF), separately. The application of PCA to multi-dimensional (S)TEM spectroscopic data sets for phase analysis (electron energy-loss and X-ray emission) has been demonstrated widely \citep{Bosman2006,Allen2011,Yaguchi2008,Parish2010}. More recently, PCA of 4D-STEM data sets for strain mapping has also been demonstrated \citep{Han2018}. PCA is deterministic and computationally fast. However, a limitation of PCA is that the orthogonal matrices comprising the output contain both positive and negative values, which can make a physical interpretation of the results challenging. In contrast, NNMF is non-deterministic requiring many iterations to converge to a solution. This results in significantly longer computation times, but a key benefit of NNMF is that the algorithm prohibits negative values and thus yields directly interpretable results \citep{Lee1999}. To date, NNMF has been employed most widely by the astrophysics community but is gaining momentum in the (S)TEM field, having been demonstrated for electron energy-loss spectrum-imaging \citep{Nicoletti2013,Ringe2017}, SPED \citep{Eggeman2015,Sunde2017,Martineau2019}, and most recently also for 4D-STEM \citep{Uesugi2020, Savitzky2020}. In this work we compare PCA and NNMF in the context of grain mapping by 4D-STEM, and discuss the characteristics and relative benefits of each classification approach. 

\section{Materials and Methods}
\label{Materials and Methods}

\subsection{4D-STEM experiments}

Samples of a fresh silica-based catalyst containing gold-palladium nanoparticles were prepared by embedding a small amount of catalyst powder with LR White resin (medium) and curing overnight at \unit[60]{$^\circ$C}. Samples were then sectioned at room temperature using a diamond knife with a Reichart Ultracut S ultramicrotome to a thickness of \unit[60]{nm}. The sections were collected on a standard copper TEM grid (300 mesh) with a lacey–carbon support (Ted Pella 01883).

The 4D-STEM experiments were performed using a double aberration-corrected modified FEI Titan 80-300 microscope (TEAM I instrument at the Molecular Foundry, LBNL), equipped with a Gatan K2 IS direct electron detector operating at 400 frames per second. The electron beam energy was \unit[300]{keV}. For grain mapping at the highest spatial resolution, an electron probe with full width at half maximum (FWHM) of \unit[0.46]{nm} was selected (convergence angle \unit[3]{mrad}, camera length \unit[230]{mm}) and a scan step size of \unit[0.25]{nm} was used. Larger fields of view were mapped using a probe of FWHM \unit[1.05]{nm}  (convergence angle \unit[1.5]{mrad}, camera length \unit[285]{mm}) implementing a scan step size of \unit[0.5]{nm}. Probe currents were \unit[$<$200]{pA}. Beam-induced damage of the nanoparticles was not observed. The largest data set comprised 32,000 diffraction patterns and surveyed an area of \unit[80$\times$100]{nm$^2$} (acquisition time \unit[80]{s}). 

For reference, high-resolution high-angle annular dark-field (HAADF) STEM images of a subset of nanoparticles were also acquired using the TEAM I instrument at \unit[300]{keV}. In addition, sample surveys using STEM-based X-ray energy-dispersive spectroscopy (XEDS) were performed at Dow Chemical using a Thermo-Fisher (FEI) Titan Themis operated at \unit[200]{keV} using Bruker AXS XFlash XEDS detectors with an energy resolution of \unit[137]{eV/channel}.  

\subsection{4D-STEM data analysis}

\subsubsection{Pre-processing}
\label{Pre-processing}

Analysis of the 4D-STEM data sets was performed in MATLAB (version R2019a). This involved a number of data pre-processing steps using custom scripts as described below, followed by grain classification by PCA and NNMF as described in Section \ref{Grain classification by PCA and NNMF}. The MATLAB scripts are available from the authors upon request.

First, the original data acquired in proprietary .dm4 (or .dm3) Gatan Digital Micrograph format was read into MATLAB giving a 4D data stack with two dimensions defining the scanned areas and two dimensions defining the diffraction patterns. The diffraction patterns were binned by a factor of two in order to reduce file size and thus speed up the subsequent computations (note, this does not affect the spatial resolution of the grain maps, since only the NBED patterns are binned). Next, the 4D stacks were converted into 3D stacks by merging the scan co-ordinates to give a number sequence of diffraction patterns.

In order to quickly visualize the nanoparticle distribution in any given surveyed area, a virtual dark-field image was reconstructed by summing the intensity values in each NBED pattern and plotting the result for each corresponding dwell point in the scan. Similarly, a quick overview of the diffraction data can be obtained by creating a single NBED pattern comprising the maximum intensity measured at each pixel (or alternatively, the mean). 

The next step is disk registration, to automatically locate all the Bragg disks in the data set. For this, an intense Bragg disk was first cropped from the NBED pattern, corresponding to the brightest pixel in the virtual dark-field image, and used to create a template. Next, we implemented a disk registration algorithm developed previously, using a hybridized standard cross-correlation and phase correlation approach, followed by Fourier upsampling \citep{Pekin2017} to determine the positions of all Bragg disks in the stack with sub-pixel precision. For data sets acquired using a beam stop to block the intense non-diffracted central disk, a mask was used to crop out bright spots emerging from the sides of the beam stop before passing the data through the disk registration algorithm so as to prevent the registration of erroneous peaks. Each Bragg disk detected was then reduced to a point, thus significantly reducing the effects of variations in disk shape and structure due to dynamical effects from the subsequent analysis. Plotting all resulting Bragg peaks in a single image, the center of the diffraction pattern was determined by computing the center of mass of opposite peak clusters to enable re-centering of the NBED stack to $(k_x,k_y)=0$. Any elliptical distortion of the diffraction pattern arising from the lens system of the microscope was then measured and corrected.

In the final pre-processing step, the NBED stack was rasterized in diffraction space to generate an output array for the subsequent PCA/NNMF routines. For the rasterization, a square mesh grid was defined of appropriate mesh size (typically one quarter of the original Bragg disk diameter) and the intensity from each Bragg peak was distributed to the nearest four corners of its corresponding mesh square, weighted according to distance. A 2D data matrix $\mathbf{X}$ was then generated, with $m$ rows corresponding to the number sequence of diffraction patterns and $n$ columns corresponding to the number sequence of rasterized NBED mesh points (essentially binned locations in diffraction space). 

\subsubsection{Grain classification by PCA and NNMF}
\label{Grain classification by PCA and NNMF}

For grain classification by PCA and NNMF, MATLAB in-built functions were used. Various custom scripts were used for plotting and analyzing the output. 

The basic principle of PCA is to reduce the dimensionality of a data set by finding a linear combination of uncorrelated variables which successively maximize statistical variance \citep{Jolliffe2016}. Typically, the input data matrix (in our case $\mathbf{X}_{m \times n}$) is first centered at the mean, allowing PCA to be performed by singular value decomposition; these two steps are the default computations performed by the MATLAB \textit{pca} function. The result of the matrix factorization can be written as
\begin{gather}\label{eq1}
    \mathbf{X}_{m \times n} 
    = \mathbf{S}_{m \times n} \mathbf{V}_{n \times n}^T,\\
    \textrm{where } \mathbf{S}_{m \times n} = \mathbf{U}_{m \times m}\,\Sigma_{m \times n}. \notag
\end{gather}
$\mathbf{U}$ and $\mathbf{V}$ are square matrices with columns and rows composed of orthogonal unit vectors, $T$ denotes a matrix transpose, and $\mathbf{\Sigma}$ is a rectangular diagonal matrix, effectively a scaling matrix. The principal components (basis vectors) are the columns of $\mathbf{S}$ listed in descending order of variance. The elements of $\mathbf{S}$ are known as the scores and the elements of $\mathbf{V}$ are known as the loadings (or coefficients). Due to the orthogonality constraint, the scores and loadings can have both positive and negative values - this can make  interpretation of the individual components challenging, since negative values are non-physical in diffraction patterns. 

Once the matrix factorization has been performed, a truncated number of components, $k$, can then be selected and used to compute a reconstructed data matrix to approximate the original, i.e. 
\begin{gather}\label{eq2}
    \mathbf{X}_{m \times n} 
    \approx \mathbf{S}_{m \times k} \mathbf{V}^T_{k \times n},\\
    k < n. \notag
\end{gather}
A common method used to guide the choice of the reduced number of components $k$ is to plot the variance (given by the squares of the eigenvalues of $\mathbf{\Sigma}$, or equivalently the eigenvalues of the covariance matrix of $\mathbf{X}$) versus the component index. This is also known as a scree plot. Generally there is first a sharp downwards trend and then the plot levels out, with the transition between these two regions representing the transition between statistically significant components and higher order components corresponding to random noise. The value of $k$ is often selected to include a subset of the first `noise' components, so as not to lose statistically significant information in the reconstruction. However, there is no general consensus on how exactly this selection should be performed, i.e.\ how far beyond the elbow the series of components should be truncated. An alternative approach is to compute the percentage of the total variance accounted for by each component. This is also called the ``percentage variance explained''. The set of components for the reconstruction can then be selected based on a predefined amount of the cumulative variance explained.

The key distinction of dimensionality reduction by NNMF is that the initial data matrix $\mathbf{X}$, which must only contain positive values, is approximated by a matrix product whose elements are also constrained to be positive \citep{Lee1999}, i.e.
\begin{gather}\label{eq3}
    \mathbf{X}_{m \times n} \approx \mathbf{W}_{m \times c} \mathbf{H}_{c \times n}, \\
    \mathbf{W}, \mathbf{H} \geq 0. \notag
\end{gather}
$\mathbf{W}$ and $\mathbf{H}$ are essentially equivalent to the $\mathbf{S}$ and $\mathbf{V}$ matrices of PCA, containing the basis vectors from the original data and their weights (also known as encoding coefficients), respectively, for each class identified. The rank $c$ gives the total number of classes (corresponding to the components of PCA) and satisfies $0<c<\min(m,n)$. The factorization proceeds iteratively, seeking to minimize the residual between $\mathbf{X}$ and $\mathbf{WH}$. In the MATLAB \textit{nnmf} function, the residual is quantified by computing the Frobenius norm (square root of the sum of the squares of the matrix elements) of $\mathbf{X-WH}$. Since the factorization may converge to a local minimum, a predefined number of repeat factorizations (known as replicates) are run with different initial values. The replicate that gives the smallest residual is selected for the final result. In our work, the number of replicates was typically set to 12. An initial number of classes for the matrix factorization, $c$, must also be selected - we chose $c$ to be at least five times greater than that expected for the data in hand, based on inspection of the scree plot obtained by PCA. A first pass of NNMF was then performed, using an alternating least-squares algorithm and random initial values for $\mathbf{W}$ and $\mathbf{H}$ (these are the default parameters in the MATLAB function). 

The number of classes $c$ was then reduced and subsequent passes of NNMF were performed as follows. First the MATLAB \textit{corr} function was used to compute a correlation matrix for $\mathbf{W}$ and $\mathbf{H}^T$, respectively, comprising the linear correlation coefficients (ranging from -1 to +1) for each pair of columns, i.e. for each class. The correlation matrices were then combined and the classes merged for the cases where the correlation factors exceeded a predefined value (typically 0.4--0.5). Finally, the merged classes were used to compute new values for $\mathbf{W}$ and $\mathbf{H}$ and NNMF was rerun using these for the initial input values together with the new value for $c$. This sequence of merging classes and re-running NNMF was repeated until the successive decrease in the number of classes leveled out to a plateau. For the data sets presented here, up to 18 passes of NNMF were performed. 

The PCA and NNMF results were visualized by reshaping the product matrices from Equations \ref{eq2} and \ref{eq3} to give image pairs for each method comprising a real-space image highlighting a particular region of the sample (i.e.\ a grain, calculated from $\mathbf{S}$ and $\mathbf{W}$) and a diffraction-space image of the corresponding pattern of Bragg peaks (calculated from $\mathbf{V}$ and $\mathbf{H}$). To aid interpretation of the peak patterns, a virtual central beam spot was added during the plotting sequence to compensate for the absence of a central disk due to the use of the beam stop during the data acquisition. The basis vectors from $\mathbf{S}$ and $\mathbf{W}$ were also used to compute a weighted average NBED pattern for each grain identified. 

Since NNMF does not use eigen-decomposition, a PCA-like scree plot of component variance versus component index cannot be generated. Thus in order to compare the ability of PCA and NNMF to capture the essence of the original data, we chose to plot the sum of squares of the residuals $\mathbf{X-SV}$ and $\mathbf{X-WH}$ versus the component/class index instead. In the case of PCA, the reconstructed matrix $\mathbf{SV}$ for increasing numbers of components can simply be computed by sequentially truncating the product matrices (which are already ordered in decreasing order of statistical importance). In contrast, in NNMF, the number of classes is one of the parameters of the factorization, thus the algorithm has to be rerun for each desired number of classes. One approach to obtain the reconstructed $\mathbf{WH}$ matrix for each NNMF class index is to run NNMF sequentially, increasing the value of $c$ each time and using the product matrices from the previous run as the starting values for the next \citep{Ren2018}. However, in our analysis, multiple passes of NNMF were already performed as part of the class merging sequence described above, each NNMF pass being for a reduced number of classes as determined using the correlation function. Hence for the NNMF portion of our matrix residual comparison, the sparse data points from the merging sequence were plotted. In addition, single passes of NNMF spanning the range of classes up to $c=100$ were also performed, and as will be seen, these data points for the single NNMF passes follow the same trend as for the NNMF merging sequence. 

\section{Results and Discussion}

\subsection{Sample overview}

\begin{figure}[t]
\centering
\includegraphics[width=0.95\linewidth]{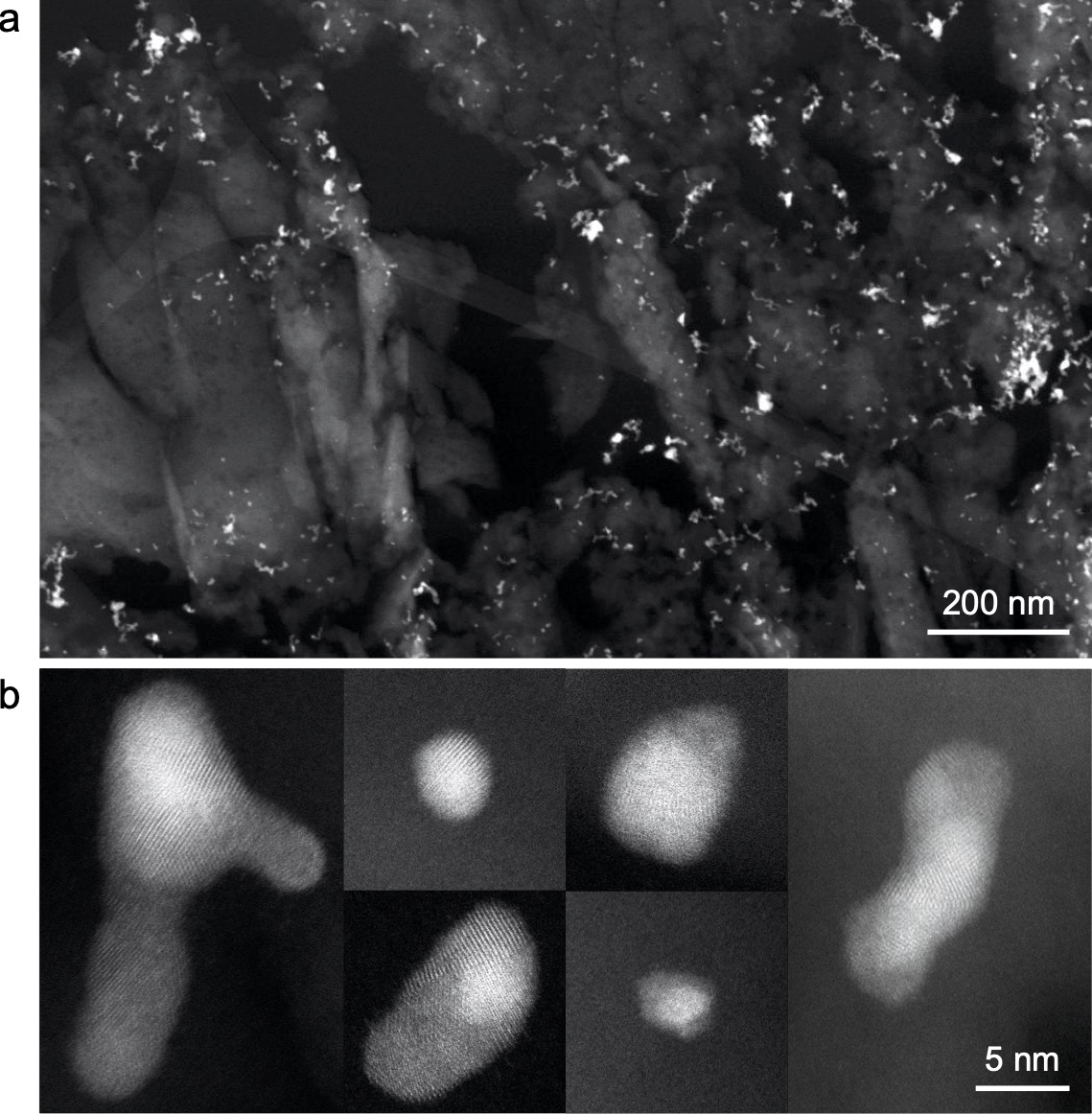}
  \caption{(a) Low-magnification HAADF-STEM of gold-palladium nanoparticles embedded in silica. (b) High-resolution HAADF-STEM of individual nanoparticles.} 
  \label{Fig1}
\end{figure}

Figure \ref{Fig1}a shows a low-magnification HAADF-STEM image of the gold-palladium nanoparticles (bright contrast) embedded in silica that were used in this study. Part of the lacey carbon TEM grid support is also visible. A set of high-resolution HAADF-STEM images of individual nanoparticles is shown in Figure \ref{Fig1}b. In the high-resolution images we see that certain regions (or grains) appear brighter than others, due to diffraction contrast dictated by the orientation of the crystal lattice planes with respect to the incident electron beam. However, the precise locations of grain boundaries within particles are challenging to discern. Elemental maps obtained by STEM-XEDS show gold- as well as palladium-rich nanoparticles, but with no clear distinction between individual grains under the electron dose and dwell times used for these experiments (see Figure S1). With these observations in mind, and given that both gold and palladium have the same crystal structure (fcc) and very similar lattice constants, it becomes clear that an alternative method to efficiently map the grains in this sample is needed, i.e.\ 4D-STEM.

\begin{figure}
\centering
\includegraphics[width=0.95\linewidth]{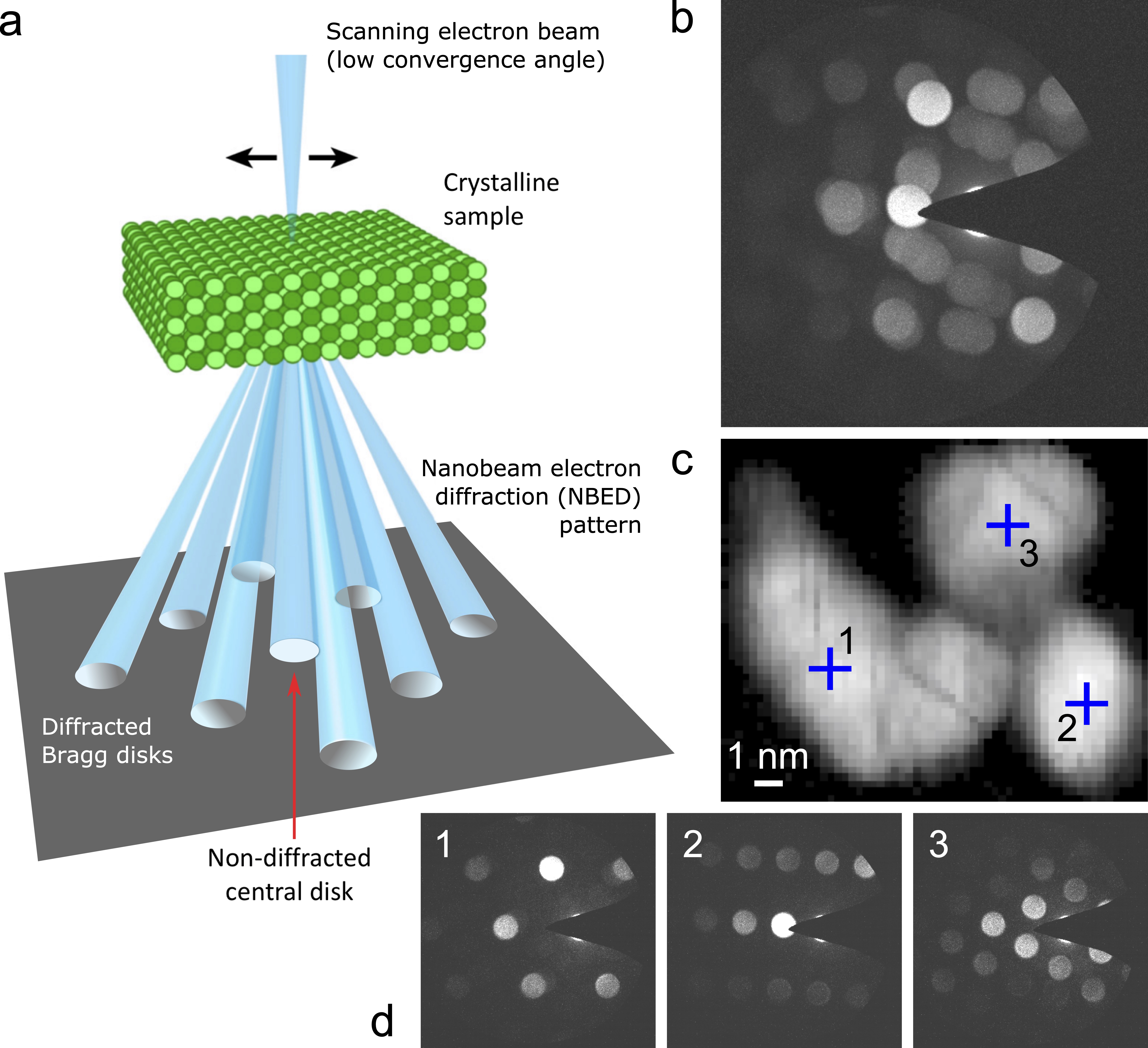}
  \caption{(a) Schematic outlining 4D-STEM data acquisition scheme used in this work. (b) Summed NBED pattern from a 4D-STEM scan of a single nanoparticle cluster, plotting the maximum intensity recorded per pixel. (c) Corresponding virtual dark-field image with pixels in three main regions marked. (d) NBED patterns for each of the three marked pixels.} 
  \label{Fig2}
\end{figure}

A schematic summarizing the 4D-STEM data acquisition method used in this study is presented in Figure \ref{Fig2}a, indicating the collection of an NBED pattern for every scan point on the sample. Plotting the maximum intensity measured per diffraction pixel across the data set gives an NBED pattern representing the entire acquisition, as shown in Figure \ref{Fig2}b. This data is for a single nanoparticle cluster scanned using a step size of \unit[0.25]{nm} over a field of view of \unit[15$\times$13]{nm$^2$} (probe FWHM \unit[0.5]{nm}). The corresponding virtual dark-field image, computed by plotting the summed NBED intensity measured per scan point, is shown in Figure \ref{Fig2}c. The three cross hairs in the virtual dark-field image mark individual scan points, with the respective NBED patterns recorded for these positions shown to the left, in Figure \ref{Fig2}d. In the following, the grain mapping analysis for this small nanoparticle cluster are presented, followed by the results for a larger field of view capturing many clusters.

\subsection{Grain classification by PCA and NNMF}

\subsubsection{Single nanoparticle cluster}

\begin{figure*}[t]
\centering
\includegraphics[width=0.95\linewidth]{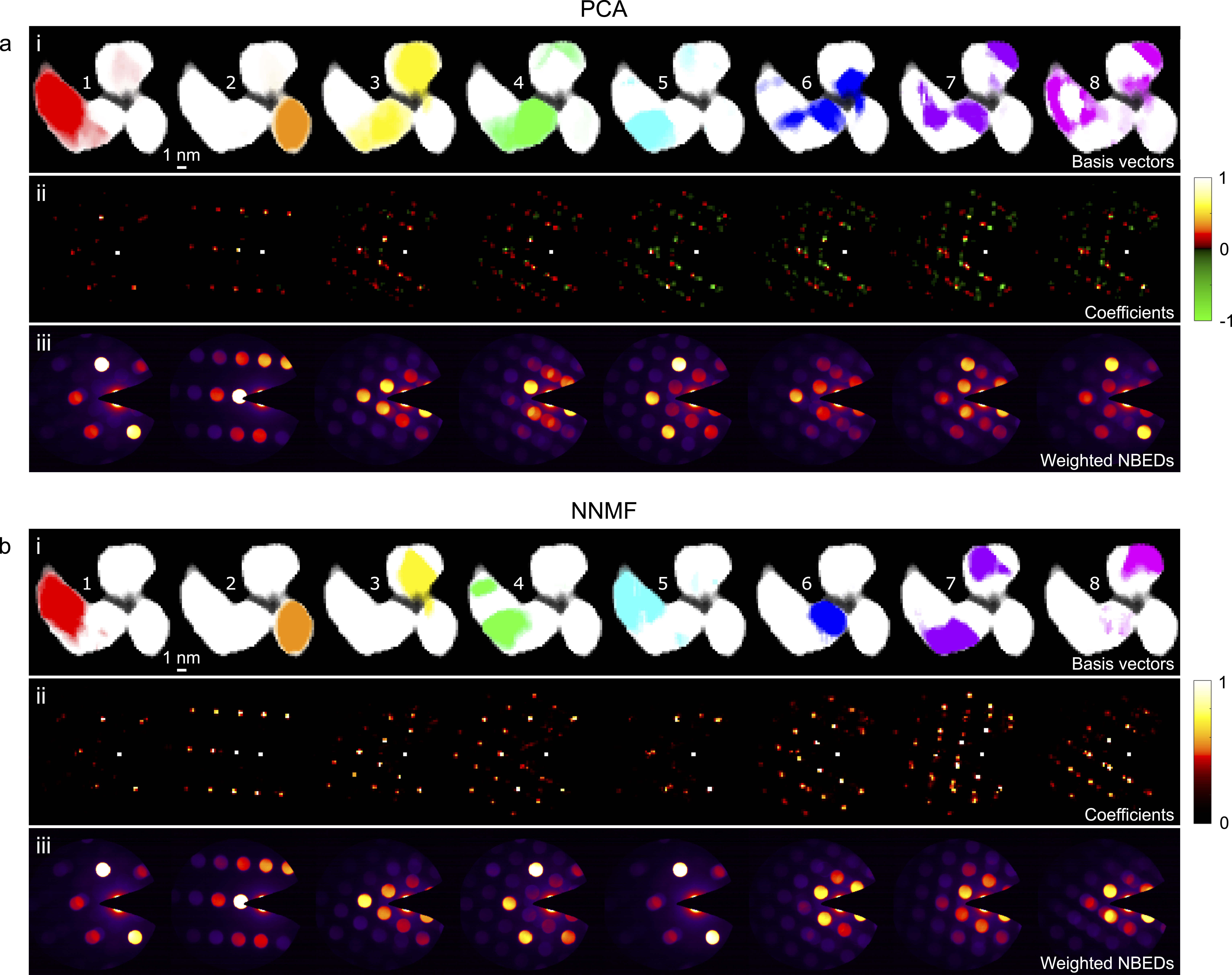}
  \caption{Grain classification results for a single nanoparticle cluster using (a) PCA and (b) NNMF to analyze the 4D-STEM data. (i) Basis vectors (re-shaped) for the first eight components/classes. (ii) Respective coefficients (re-shaped) with color bars indicating the intensity values. (iii) Weighted NBED patterns reconstructed for each component/class.}
  \label{Fig3}
\end{figure*}

PCA and NNMF grain classification results for the single nanoparticle cluster are shown in Figures \ref{Fig3}a and \ref{Fig3}b, respectively, obtained following the data pre-processing and matrix factorization procedures described in Section 2.2. The top row (i) of each subfigure shows the basis vectors (re-shaped) for the first eight classes identified, plotted in order of decreasing statistical significance. These images map the spatial regions of the nanoparticle corresponding to each class. The second row (ii) shows the respective coefficients (re-shaped) representing the diffraction `fingerprints' obtained for each class. The Bragg peaks appear as points rather than the original disks, since the disks were reduced to points in the disk registration pre-processing step to enable grain mapping without influence from dynamical effects. To aid interpretation of the diffraction patterns, which were collected using a beam stop to mask the intense central beam, a virtual central beam spot has been added to these images. Finally, in the third row (iii), the reconstructed NBED pattern corresponding to each component/class is plotted, obtained using the basis vectors to compute weighted sums from the NBED patterns in the original 4D-STEM data stack. We show the results for a total of 8 components/classes, since this was the number of classes obtained after the NNMF class merging sequence (the initial number of classes had been set to 16). Scree plot analysis of the PCA results (Figure S2) also indicates that 8 components is a reasonable number.

The color bars inserted alongside the coefficients in Figure \ref{Fig3} show the range of intensity values that the coefficients possess. A comparison of these ranges for PCA (Figure \ref{Fig3}a) and NNMF (Figure \ref{Fig3}b) highlights one of the key distinguishing features of the two factorization techniques, i.e.\ that the output from PCA can have both positive and negative values, whereas in the case of NNMF, only positive values are allowed. In this example, we see that the negative (green) peaks in the PCA coefficients of Figure \ref{Fig3}a row (ii) increasingly show up for the third component onwards. Negative output is also obtained for the PCA basis vectors, but for simplicity, only the positive values were used to create the PCA basis vector maps shown in the figure. The occurrence of both positive and negative values in PCA is a direct consequence of the orthogonal factorization constraint, which can yield non-physical results. That is, in diffraction data and in imaging in general, the physical measurement values are by nature all positive, hence interpretation of negative coefficients (in our case, negative Bragg peaks) is challenging. However, the weighted NBED patterns in Figure \ref{Fig3}a row (iii) show diffraction patterns that are all `real', with positive peaks, since these were constructed by weighting the original NBED data.

A further direct consequence of the orthogonal factorization of PCA is the tendency to produce mixed phases in the output (generally for the higher order components), since in reality phases do not necessarily have to be linearly uncorrelated. For example, the weighted NBED pattern for the fourth PCA component in Figure \ref{Fig3}a row (iii) clearly comprises two patterns indexing to the same lattice reflection with a slight rotational offset. It is worth noting that phase mixing in our results can also be real, resulting from overlapping grains in the beam direction. Indeed, both PCA and NNMF are sensitive to detecting mixed phases if present, since the reconstructed data is represented as a superposition of basis vectors, as for example shown in the analysis of various spectral data sets \citep{Long2009,Kannan2018} and of SPED data sets \citep{Sunde2017}. However, in our example, the weighted NBED patterns from the NNMF analysis do not indicate overlapping phases to the same extent as is seen in the PCA results. Hence the phase mixing observed in the PCA output would appear to largely be an artifact from the orthogonality-constrained PCA factorization. 

The NNMF results shown in Figure \ref{Fig3}b give a similar classification of grains to the PCA results, but there are some important distinctions. Firstly, the NNMF coefficients (and basis vectors) by definition contain positive values only, as reflected by the color bar in row (ii), i.e.\ no negative peaks. Thus each re-shaped coefficient matrix represents a real diffraction pattern. Secondly, upon inspection of the NNMF coefficients and weighted NBED patterns we see that they are largely discrete for each class, with satellite peaks and overlapping disks less prevalent than in the PCA results. This highlights the ability of NNMF to identify discrete components with less tendency to capture mixed phases in a single class. We also note that the weighted NBED patterns for the first and fifth classes corresponding to the left hand side of the cluster are very similar, suggesting that there may be two grains overlapping in depth with only a slight misorientation between them. Rather than classifying these as a mixed phase, NNMF discerned them as separate phases. 

In terms of computation time (using a 2017 MacBook Pro Intel Core i5, \unit[8]{GB} RAM), the PCA algorithm took \unit[4]{minutes} to complete, whereas the NNMF algorithm, including the class merging sequence (8 merges) took \unit[40]{minutes}. The size of the input 4D-STEM data set was \unit[1]{GB}.

Finally, we found that the NNMF classification was much more sensitive to the mesh spacing chosen to rasterize the diffraction patterns in the data pre-processing step described in Section 2.2.1. For example, mesh spacings of 5 or 6 pixels consistently yielded NNMF results in which the first (highest variance) component was the grain on the left hand side of the cluster (i.e.\ the red grain in Figures \ref{Fig3}a and b), whereas using a mesh spacing of 7 or 8, the order in the NNMF output changed and the orange grain became the most statistically important. In contrast, in the PCA case, the order of components was not affected by these changes. This again points to the enhanced sensitivity of NNMF to subtle differences in the input data set. For the data analysis presented here, a rasterization mesh size of 6 pixels was used in all cases.

\subsubsection{Distribution of nanoparticles}

\begin{figure}
\centering
\includegraphics[width=0.95\linewidth]{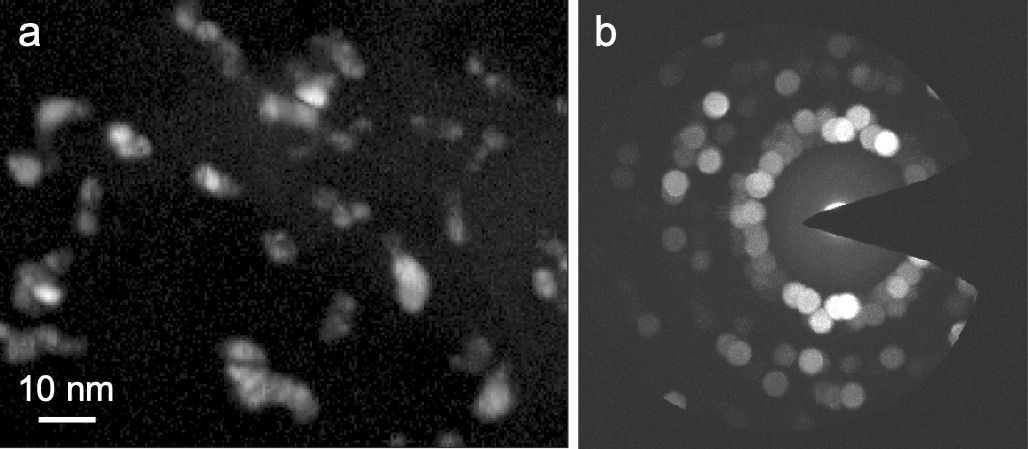}
  \caption{(a) Virtual dark-field image for a distribution of nanoparticles surveyed in a 4D-STEM scan. (b) Summed NBED pattern plotting the maximum intensity recorded per pixel.}
  \label{Fig4}
\end{figure}

\begin{figure*}[t]
\centering
\includegraphics[width=0.95\linewidth]{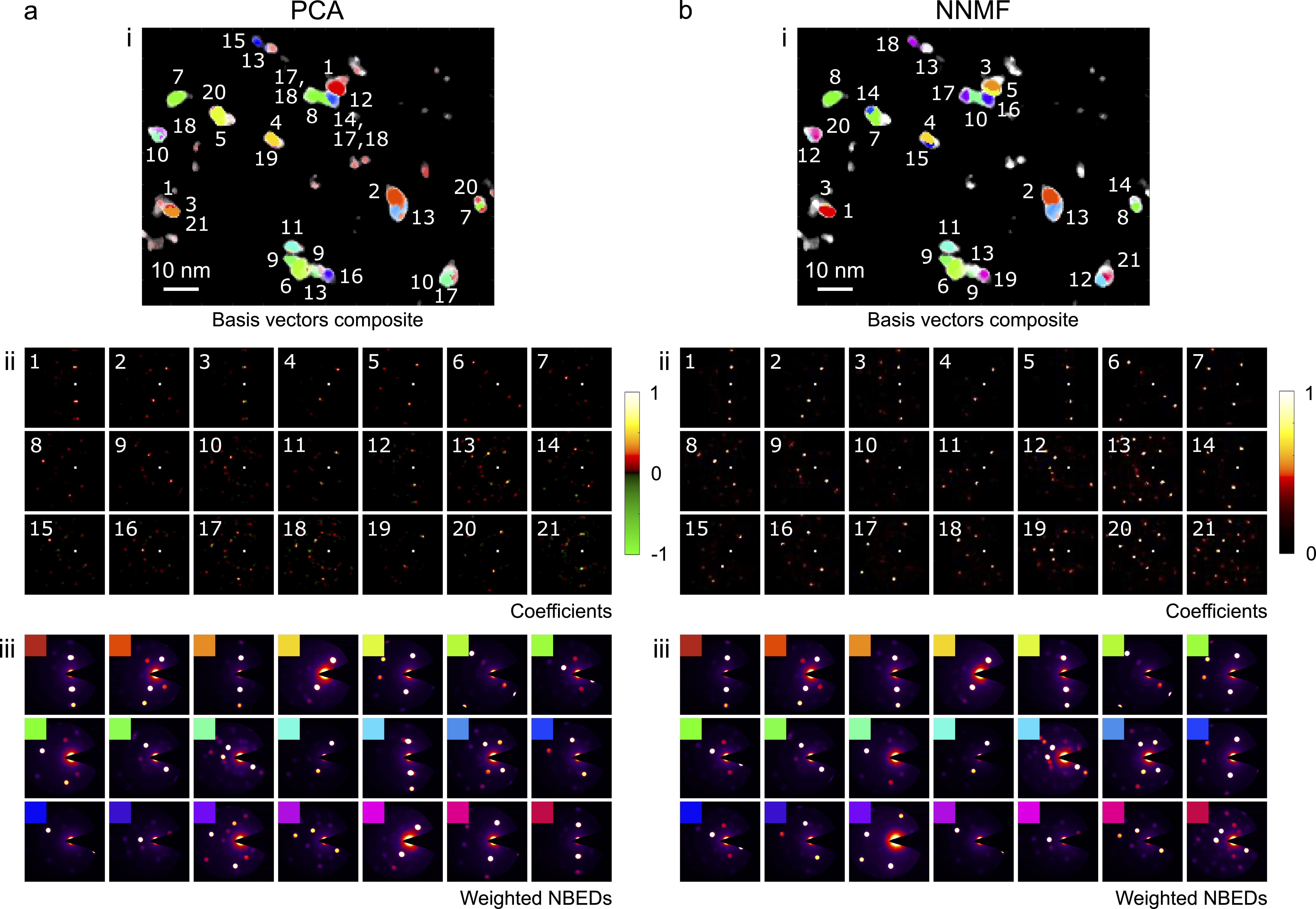}
  \caption{Grain classification results for a distribution of nanoparticles using (a) PCA and (b) NNMF to analyze the 4D-STEM data. (i) Composite maps of the basis vectors obtained (numbered), (ii) coefficients for each component/class with color bars indicating the intensity values, (iii) weighted NBED patterns reconstructed and color-coded for each component/class.}
  \label{Fig5}
\end{figure*}

The virtual dark-field image and corresponding 4D-STEM NBED stack for the region of the sample surveyed over a larger field of view (\unit[80$\times$100]{nm$^2$}) are shown in Figures \ref{Fig4}a and \ref{Fig4}b, respectively. For this data set, a scan step size of \unit[0.5]{nm} and probe size of \unit[1]{nm} (FWHM) were selected. The PCA and NNMF grain classification results are presented in Figure \ref{Fig5}, showing composite maps of the (numbered) basis vectors in a(i) and b(i), the individual coefficients for each component/class in a(ii) and b(ii), and the weighted NBED patterns in a(iii) and b(iii). The number of components/classes plotted is 21, since this was the number of classes obtained after the NNMF class merging sequence (the initial number of classes had been set to 100). Scree plot analysis of the PCA results (Figure S3) also indicates that \unit[$\sim$]{20} is a reasonable estimate for the number of distinct phases present in the surveyed area.

Comparing the PCA and NNMF results of Figure \ref{Fig5} it can be seen that essentially the grain maps in each case are very similar. There is some variation in the number order of the grains, and in the PCA case, there is a certain degree of phase mixing (most evident in the composite map, where the labels for the overlapping grains are comma-separated). These differences reflect those previously discussed in the PCA/NNMF analysis of the small nanoparticle cluster.

\begin{figure}
\centering
\includegraphics[width=0.95\linewidth]{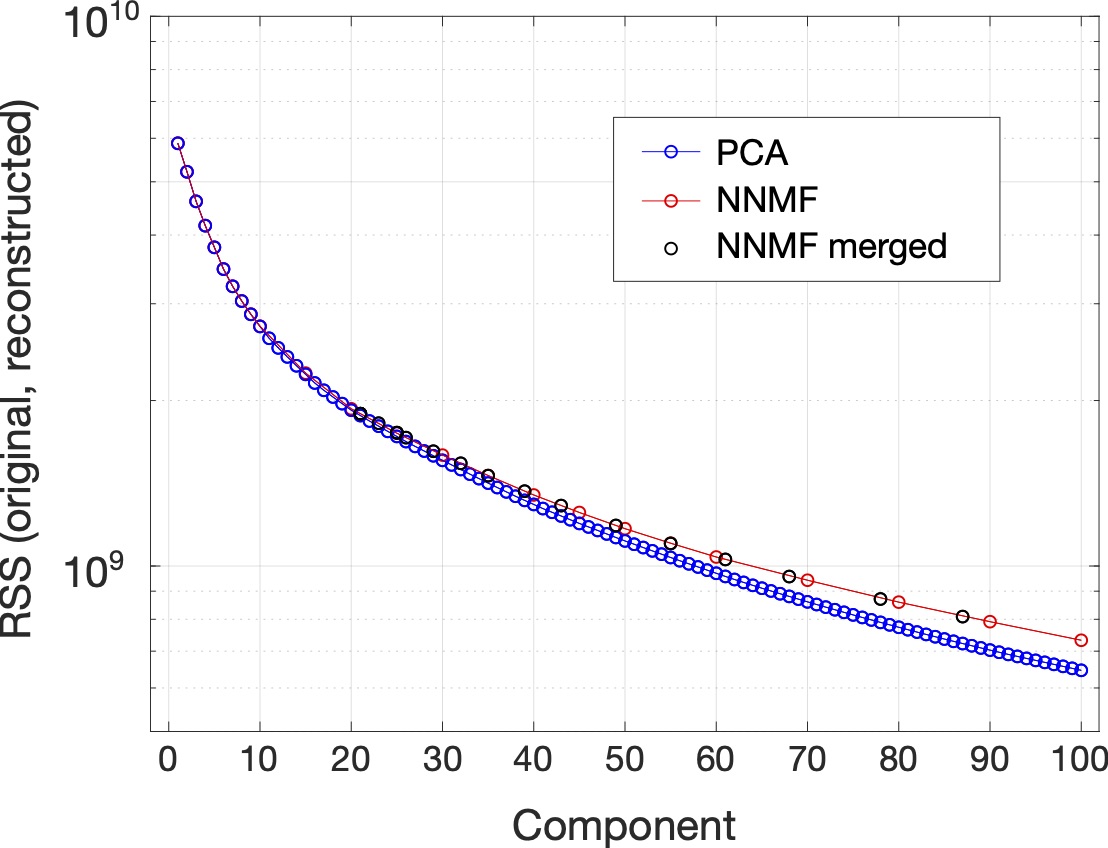}
  \caption{Matrix residual-based comparison of the PCA and NNMF results for the nanoparticle distribution: Plot of the squared sums of the matrix elements of the residual matrices $\mathbf{X-SV}$ (PCA) and $\mathbf{X-WH}$ (NNMF) versus component/class index. Data points for PCA in blue, for NNMF without merging in red, and for NNMF after merging (starting with $c=100$) in black.} 
  \label{Fig6}
\end{figure}

In order to further compare the PCA and NNMF results, plots of the sum of squares of the residual matrices (i.e.\ (initial data matrix) $-$ (reconstructed data matrix)) versus component/class index have been generated, as shown in Figure \ref{Fig6} and discussed previously at the end of Section \ref{Grain classification by PCA and NNMF}. The PCA data points (blue circles) have been obtained by sequentially truncating the reconstructed PCA matrix in increments of one component, whereas the data points for NNMF have been obtained from individual passes of NNMF using a different number of input classes \textit{without} subsequent merging (red circles), and also using the $c=100$ data set and plotting the data points corresponding to each class merging step (black circles). As the number of components increases, the matrix residuals rapidly decrease. This confirms that a larger number of components allows the PCA/NNMF reconstructions to capture the original data more effectively. Furthermore, as the NNMF classes are merged (moving from right to left in the plot), the residuals gradually increase. This is because each merging step introduces a small error (by definition the classes being merged were not perfect matches, but rather satisfied a pre-defined correlation threshold).

Up to $c\approx 20$, the PCA and NNMF traces in Figure \ref{Fig6} essentially overlap, indicating close similarity between the reconstructions obtained using each technique up to this point (as also indicated by the comparison of PCA and NNMF output shown in Figure \ref{Fig5}). 
Then for $c>20$, where components start to represent noise, the two traces start to deviate with the PCA trace progressively reaching lower residual values than the NNMF trace. This can be explained by the tendency of PCA to overfit the data compared to NNMF \citep{Ren2018}.

Due to the file size of the large field-of-view data set (\unit[67]{GB}), these computations were performed using a workstation with two Intel Xenon Silver 4114 CPUs with \unit[512]{GB} RAM. For the PCA algorithm the computation time was \unit[55]{s}, whereas for the NNMF algorithm, including the class merging sequence (16 merges) it was close to \unit[44]{hours}. In order to speed up the NNMF computation, the size of the input data set could be reduced by using more binning and/or by using thresholding to create a 2D probability distribution of all grains and then clustering the detected Bragg disks, for example using Voronoi cells \citep{Savitzky2020}.

\section{Conclusions}

We have shown that 4D-STEM using a sub-nanometer probe size with a beam overlap of \unit[50]{\%} combined with fast direct electron detection enables efficient grain mapping at high spatial resolution. Since relatively large fields of view (by STEM standards) can be probed in seconds to minutes, high-resolution mapping while maintaining minimal sample drift can be achieved. Beam and sample alignments for the 4D-STEM method used here are also greatly simplified compared with precession-based methods. 

Developments in direct electron detector technology continue to transform the field, with data acquisition speeds greater than ten thousand frames per second now realized \citep{Nord2020,Ercius2020}, i.e.\ two orders of magnitude faster than the direct-electron detector used in this work. With this significant advance, 4D-STEM is now poised to enable efficient grain mapping over fields of view approaching the sizes surveyed by EBSD in the SEM, but with the spatial resolution provided by STEM. Thus, fast high-resolution grain mapping over large fields of view for high-throughput sub-nanometer grain analysis becomes possible. For the analysis of nanoparticle catalysts such as those surveyed in this work, this paves the way towards statistical analysis of catalyst grain size distributions and orientations through the screening of thousands of particles before and after use, providing crucial insight into the factors affecting catalyst efficiency and lifetime. Future work can also make use of multibeam electron diffraction, using an aperture to form multiple probes that increase the angular range surveyed per scan point and thus enable robust indexing due to the richer diffraction patterns obtained \citep{Hong2020}.

Ever increasing data acquisition speeds mean ever increasing file sizes. Consequently, high levels of automation for parameter optimization in the data analysis routines will be essential. In the present work, we have demonstrated several critical semi-automated data pre-processing routines and shown how parameter choice can affect the final classification results. In the future, various open-source software routines that are currently being developed will be well-positioned to increase automation for enhanced multiparameter optimization \citep{Savitzky2019,Clausen2020,Johnstone2021}. Further benefit can be drawn from the use of patterned probes generated using a binary mask in the probe-forming aperture in order to facilitate the subsequent disk registration \citep{Zeltmann2020}. Finally, we have shown that feature learning using PCA offers fast classification of phases and is thus well-suited for a first pass analysis. NNMF computation is much more time-intensive, but offers direct interpretation of the matrix output due to the non-negativity constraint, delivers results free from orthogonality-induced phase-mixing, and is also more sensitive to subtle variations in the input data.

\section*{Acknowledgements}

F.I.A., S.J.R., G.F.M. and A.M.M. gratefully acknowledge support from the Dow University Partnership Initiative Program. Work at the Molecular Foundry at Lawrence Berkeley National Laboratory was supported by the U.S. Department of Energy under Contract \# DE-AC02-05CH11231.  J.C. acknowledges additional support from the Presidential Early Career Award for Scientists and Engineers (PECASE) through the U.S. Department of Energy.    

\newpage

\bibliographystyle{elsarticle-harv}

\bibliography{bib.bib}

\begin{thebibliography}{35}
\expandafter\ifx\csname natexlab\endcsname\relax\def\natexlab#1{#1}\fi
\providecommand{\url}[1]{\texttt{#1}}
\providecommand{\href}[2]{#2}
\providecommand{\path}[1]{#1}
\providecommand{\DOIprefix}{doi:}
\providecommand{\ArXivprefix}{arXiv:}
\providecommand{\URLprefix}{URL: }
\providecommand{\Pubmedprefix}{pmid:}
\providecommand{\doi}[1]{\href{http://dx.doi.org/#1}{\path{#1}}}
\providecommand{\Pubmed}[1]{\href{pmid:#1}{\path{#1}}}
\providecommand{\bibinfo}[2]{#2}
\ifx\xfnm\relax \def\xfnm[#1]{\unskip,\space#1}\fi
\bibitem[{Allen et~al.(2011)Allen, Watanabe, Lee, Balsara and
  Minor}]{Allen2011}
\bibinfo{author}{Allen, F.I.}, \bibinfo{author}{Watanabe, M.},
  \bibinfo{author}{Lee, Z.}, \bibinfo{author}{Balsara, N.P.},
  \bibinfo{author}{Minor, A.M.}, \bibinfo{year}{2011}.
\newblock \bibinfo{title}{Chemical mapping of a block copolymer electrolyte by
  low-loss {EFTEM} spectrum-imaging and principal component analysis}.
\newblock \bibinfo{journal}{Ultramicroscopy} \bibinfo{volume}{111},
  \bibinfo{pages}{239}.
\newblock \DOIprefix\doi{10.1016/j.ultramic.2010.11.035}.
\bibitem[{B{\'e}ch{\'e} et~al.(2009)B{\'e}ch{\'e}, Rouvi{\`e}re, Cl{\'e}ment
  and Hartmann}]{Beche2009}
\bibinfo{author}{B{\'e}ch{\'e}, A.}, \bibinfo{author}{Rouvi{\`e}re, J.L.},
  \bibinfo{author}{Cl{\'e}ment, L.}, \bibinfo{author}{Hartmann, J.M.},
  \bibinfo{year}{2009}.
\newblock \bibinfo{title}{Improved precision in strain measurement using
  nanobeam electron diffraction}.
\newblock \bibinfo{journal}{Appl. Phys. Lett.} \bibinfo{volume}{95},
  \bibinfo{pages}{123114}.
\newblock \DOIprefix\doi{10.1063/1.3224886}.
\bibitem[{Bosman et~al.(2006)Bosman, Watanabe, Alexandar and
  Keast}]{Bosman2006}
\bibinfo{author}{Bosman, M.}, \bibinfo{author}{Watanabe, M.},
  \bibinfo{author}{Alexandar, D.T.L.}, \bibinfo{author}{Keast, V.J.},
  \bibinfo{year}{2006}.
\newblock \bibinfo{title}{Mapping chemical and bonding information using
  multivariate analysis of electron energy-loss spectrum images}.
\newblock \bibinfo{journal}{Ultramicroscopy} \bibinfo{volume}{106},
  \bibinfo{pages}{1024}.
\newblock \DOIprefix\doi{10.1016/j.ultramic.2006.04.016}.
\bibitem[{Clausen et~al.(2020)Clausen, Weber, Ruzaeva, Migunov, Baburajan,
  Bahuleyan, Caron, Chandra, Halder, Nord, M{\"u}ller-Caspary and
  Dunin-Borkowski}]{Clausen2020}
\bibinfo{author}{Clausen, A.}, \bibinfo{author}{Weber, D.},
  \bibinfo{author}{Ruzaeva, K.}, \bibinfo{author}{Migunov, V.},
  \bibinfo{author}{Baburajan, A.}, \bibinfo{author}{Bahuleyan, A.},
  \bibinfo{author}{Caron, J.}, \bibinfo{author}{Chandra, R.},
  \bibinfo{author}{Halder, S.}, \bibinfo{author}{Nord, M.},
  \bibinfo{author}{M{\"u}ller-Caspary, K.}, \bibinfo{author}{Dunin-Borkowski,
  R.E.}, \bibinfo{year}{2020}.
\newblock \bibinfo{title}{{LiberTEM}: Software platform for scalable
  multidimensional data processing in transmission electron microscopy}.
\newblock \bibinfo{journal}{Journal of Open Source Software}
  \bibinfo{volume}{5}, \bibinfo{pages}{2006}.
\newblock \DOIprefix\doi{10.21105/joss.02006}.
\bibitem[{Eggeman et~al.(2015)Eggeman, Krakow and Midgley}]{Eggeman2015}
\bibinfo{author}{Eggeman, A.S.}, \bibinfo{author}{Krakow, R.},
  \bibinfo{author}{Midgley, P.A.}, \bibinfo{year}{2015}.
\newblock \bibinfo{title}{Scanning precession electron tomography for
  three-dimensional nanoscale orientation imaging and crystallographic
  analysis}.
\newblock \bibinfo{journal}{Nature Communications} \bibinfo{volume}{6},
  \bibinfo{pages}{7267}.
\newblock \DOIprefix\doi{10.1038/ncomms8267}.
\bibitem[{Ercius et~al.(2020)Ercius, Johnson, Brown, Pelz, Hsu, Draney, Fong,
  Goldschmidt, Joseph, Lee, Ciston, Ophus, Scott, Paul, Hanwell, Harris, Avery,
  Stezelberger, Tindall, Ramesh, Minor and Denes}]{Ercius2020}
\bibinfo{author}{Ercius, P.}, \bibinfo{author}{Johnson, I.},
  \bibinfo{author}{Brown, H.}, \bibinfo{author}{Pelz, P.},
  \bibinfo{author}{Hsu, S.L.}, \bibinfo{author}{Draney, B.},
  \bibinfo{author}{Fong, E.}, \bibinfo{author}{Goldschmidt, A.},
  \bibinfo{author}{Joseph, J.}, \bibinfo{author}{Lee, J.},
  \bibinfo{author}{Ciston, J.}, \bibinfo{author}{Ophus, C.},
  \bibinfo{author}{Scott, M.}, \bibinfo{author}{Paul, D.},
  \bibinfo{author}{Hanwell, M.}, \bibinfo{author}{Harris, C.},
  \bibinfo{author}{Avery, P.}, \bibinfo{author}{Stezelberger, T.},
  \bibinfo{author}{Tindall, C.}, \bibinfo{author}{Ramesh, R.},
  \bibinfo{author}{Minor, A.}, \bibinfo{author}{Denes, P.},
  \bibinfo{year}{2020}.
\newblock \bibinfo{title}{The 4{D} camera--an 87 k{H}z frame-rate detector for
  counted 4{D}-{STEM} experiments}.
\newblock \bibinfo{journal}{Microscopy and Microanalysis} \bibinfo{volume}{26},
  \bibinfo{pages}{1896--1897}.
\newblock \DOIprefix\doi{10.1017/S1431927620019753}.
\bibitem[{Fang et~al.(2019)Fang, Wen, Allen, Ophus, Han, Kirkland, Kaxiras and
  Warner}]{Fang2019}
\bibinfo{author}{Fang, S.}, \bibinfo{author}{Wen, Y.}, \bibinfo{author}{Allen,
  C.S.}, \bibinfo{author}{Ophus, C.}, \bibinfo{author}{Han, G.G.D.},
  \bibinfo{author}{Kirkland, A.I.}, \bibinfo{author}{Kaxiras, E.},
  \bibinfo{author}{Warner, J.H.}, \bibinfo{year}{2019}.
\newblock \bibinfo{title}{Atomic electrostatic maps of 1{D} channels in 2{D}
  semiconductors using 4{D} scanning transmission electron microscopy}.
\newblock \bibinfo{journal}{Nature Communications} \bibinfo{volume}{10},
  \bibinfo{pages}{1127}.
\newblock \DOIprefix\doi{10.1038/s41467-019-08904-9}.
\bibitem[{Gammer et~al.(2015)Gammer, Ozdol, Liebscher and Minor}]{Gammer2015}
\bibinfo{author}{Gammer, C.}, \bibinfo{author}{Ozdol, V.B.},
  \bibinfo{author}{Liebscher, C.H.}, \bibinfo{author}{Minor, A.M.},
  \bibinfo{year}{2015}.
\newblock \bibinfo{title}{Diffraction contrast imaging using virtual
  apertures}.
\newblock \bibinfo{journal}{Ultramicroscopy} \bibinfo{volume}{155},
  \bibinfo{pages}{1}.
\newblock \DOIprefix\doi{10.1016/j.ultramic.2015.03.015}.
\bibitem[{Han et~al.(2018)Han, Nguyen, Cao, Cueva, Xie, Tate, Purohit, Gruner,
  Park and Muller}]{Han2018}
\bibinfo{author}{Han, Y.}, \bibinfo{author}{Nguyen, K.}, \bibinfo{author}{Cao,
  M.}, \bibinfo{author}{Cueva, P.}, \bibinfo{author}{Xie, S.},
  \bibinfo{author}{Tate, M.W.}, \bibinfo{author}{Purohit, P.},
  \bibinfo{author}{Gruner, S.M.}, \bibinfo{author}{Park, J.},
  \bibinfo{author}{Muller, D.A.}, \bibinfo{year}{2018}.
\newblock \bibinfo{title}{Strain mapping of two-dimensional heterostructures
  with subpicometer precision}.
\newblock \bibinfo{journal}{Nano Letters} \bibinfo{volume}{18},
  \bibinfo{pages}{3746}.
\newblock \DOIprefix\doi{10.1021/acs.nanolett.8b00952}.
\bibitem[{Hong et~al.(2020)Hong, Zeltmann, Savitzky, Rangel~DaCosta,
  M{\"u}ller, Minor, Bustillo and Ophus}]{Hong2020}
\bibinfo{author}{Hong, X.}, \bibinfo{author}{Zeltmann, S.E.},
  \bibinfo{author}{Savitzky, B.H.}, \bibinfo{author}{Rangel~DaCosta, L.},
  \bibinfo{author}{M{\"u}ller, A.}, \bibinfo{author}{Minor, A.M.},
  \bibinfo{author}{Bustillo, K.C.}, \bibinfo{author}{Ophus, C.},
  \bibinfo{year}{2020}.
\newblock \bibinfo{title}{Multibeam electron diffraction}.
\newblock \bibinfo{journal}{Microsc. Microanal.} ,
  \bibinfo{pages}{1--11}\DOIprefix\doi{10.1017/S1431927620024770}.
\bibitem[{Jiang et~al.(2018)Jiang, Chen, Han, Deb, Gao, Xie, Purohit, Tate,
  Park, Gruner, Elser and Muller}]{Jiang2018}
\bibinfo{author}{Jiang, Y.}, \bibinfo{author}{Chen, Z.}, \bibinfo{author}{Han,
  Y.}, \bibinfo{author}{Deb, P.}, \bibinfo{author}{Gao, H.},
  \bibinfo{author}{Xie, S.}, \bibinfo{author}{Purohit, P.},
  \bibinfo{author}{Tate, M.W.}, \bibinfo{author}{Park, J.},
  \bibinfo{author}{Gruner, S.M.}, \bibinfo{author}{Elser, V.},
  \bibinfo{author}{Muller, D.A.}, \bibinfo{year}{2018}.
\newblock \bibinfo{title}{Electron ptychography of 2{D} materials to deep
  sub-{\aa}ngstr{\"o}m resolution}.
\newblock \bibinfo{journal}{Nature} \bibinfo{volume}{559},
  \bibinfo{pages}{343}.
\newblock \DOIprefix\doi{10.1038/s41586-018-0298-5}.
\bibitem[{Johnstone et~al.(2021)Johnstone, Crout, Nord, Laulainen, Høgås,
  Opheim, Martineau, Francis, Bergh, Prestat, Smeets, Ross, Collins, Hjorth,
  Danaie, Furnival, Jannis, Cautaerts, Jacobsen, Herzing, Poon, Ånes, Morzy, ,
  Doherty, Iqbal, Ostasevicius, mvonlany and Tovey}]{Johnstone2021}
\bibinfo{author}{Johnstone, D.N.}, \bibinfo{author}{Crout, P.},
  \bibinfo{author}{Nord, M.}, \bibinfo{author}{Laulainen, J.},
  \bibinfo{author}{Høgås, S.}, \bibinfo{author}{Opheim, E.},
  \bibinfo{author}{Martineau, B.}, \bibinfo{author}{Francis, C.},
  \bibinfo{author}{Bergh, T.}, \bibinfo{author}{Prestat, E.},
  \bibinfo{author}{Smeets, S.}, \bibinfo{author}{Ross, A.},
  \bibinfo{author}{Collins, S.}, \bibinfo{author}{Hjorth, I.},
  \bibinfo{author}{Danaie, M.}, \bibinfo{author}{Furnival, T.},
  \bibinfo{author}{Jannis, D.}, \bibinfo{author}{Cautaerts, N.},
  \bibinfo{author}{Jacobsen, E.}, \bibinfo{author}{Herzing, A.},
  \bibinfo{author}{Poon, T.}, \bibinfo{author}{Ånes, H.W.},
  \bibinfo{author}{Morzy, J.}, , \bibinfo{author}{Doherty, T.},
  \bibinfo{author}{Iqbal, A.}, \bibinfo{author}{Ostasevicius, T.},
  \bibinfo{author}{mvonlany}, \bibinfo{author}{Tovey, R.},
  \bibinfo{year}{2021}.
\newblock \bibinfo{title}{pyxem 0.13.0}.
\newblock \DOIprefix\doi{10.5281/zenodo.4436723}.
\bibitem[{Jolliffe and Cadima(2016)}]{Jolliffe2016}
\bibinfo{author}{Jolliffe, I.T.}, \bibinfo{author}{Cadima, J.},
  \bibinfo{year}{2016}.
\newblock \bibinfo{title}{Principal component analysis: a review and recent
  developments}.
\newblock \bibinfo{journal}{Phil. Trans. R. Soc. A} \bibinfo{volume}{374},
  \bibinfo{pages}{20150202}.
\newblock \DOIprefix\doi{10.1098/rsta.2015.0202}.
\bibitem[{Kannan et~al.(2018)Kannan, Ievlev, Laanait, Ziatdinov, Vasudevan,
  Jesse and Kalinin}]{Kannan2018}
\bibinfo{author}{Kannan, R.}, \bibinfo{author}{Ievlev, A.V.},
  \bibinfo{author}{Laanait, N.}, \bibinfo{author}{Ziatdinov, M.A.},
  \bibinfo{author}{Vasudevan, R.K.}, \bibinfo{author}{Jesse, S.},
  \bibinfo{author}{Kalinin, S.V.}, \bibinfo{year}{2018}.
\newblock \bibinfo{title}{Deep data analysis via physically constrained linear
  unmixing: universal framework, domain examples, and a community-wide
  platform}.
\newblock \bibinfo{journal}{Adv Struct Chem Imaging} \bibinfo{volume}{4},
  \bibinfo{pages}{6}.
\newblock \DOIprefix\doi{10.1186/s40679-018-0055-8}.
\bibitem[{Keller and Geiss(2012)}]{Keller2012}
\bibinfo{author}{Keller, R.R.}, \bibinfo{author}{Geiss, R.H.},
  \bibinfo{year}{2012}.
\newblock \bibinfo{title}{Transmission {EBSD} from 10 nm domains in a scanning
  electron microscope}.
\newblock \bibinfo{journal}{Journal of Microscopy} \bibinfo{volume}{245},
  \bibinfo{pages}{245}.
\newblock \DOIprefix\doi{10.1111/j.1365-2818.2011.03566.x}.
\bibitem[{Lee and Seung(1999)}]{Lee1999}
\bibinfo{author}{Lee, D.D.}, \bibinfo{author}{Seung, H.S.},
  \bibinfo{year}{1999}.
\newblock \bibinfo{title}{Learning the parts of objects by non-negative matrix
  factorization}.
\newblock \bibinfo{journal}{Nature} \bibinfo{volume}{401},
  \bibinfo{pages}{788}.
\newblock \DOIprefix\doi{10.1038/44565}.
\bibitem[{Long et~al.(2009)Long, Bunker, Li, Karen and Takeuchi}]{Long2009}
\bibinfo{author}{Long, C.J.}, \bibinfo{author}{Bunker, D.},
  \bibinfo{author}{Li, X.}, \bibinfo{author}{Karen, V.L.},
  \bibinfo{author}{Takeuchi, I.}, \bibinfo{year}{2009}.
\newblock \bibinfo{title}{Rapid identification of structural phases in
  combinatorial thin-film libraries using x-ray diffraction and non-negative
  matrix factorization}.
\newblock \bibinfo{journal}{Review of Scientific Instruments}
  \bibinfo{volume}{80}, \bibinfo{pages}{103902}.
\newblock \DOIprefix\doi{10.1063/1.3216809}.
\bibitem[{Martineau et~al.(2019)Martineau, Johnstone, van Helvoort, Midgley and
  Eggeman}]{Martineau2019}
\bibinfo{author}{Martineau, B.H.}, \bibinfo{author}{Johnstone, D.N.},
  \bibinfo{author}{van Helvoort, A.T.J.}, \bibinfo{author}{Midgley, P.A.},
  \bibinfo{author}{Eggeman, A.S.}, \bibinfo{year}{2019}.
\newblock \bibinfo{title}{Unsupervised machine learning applied to scanning
  precession electron diffraction data}.
\newblock \bibinfo{journal}{Advanced Structural and Chemical Imaging}
  \bibinfo{volume}{5}, \bibinfo{pages}{3}.
\newblock \DOIprefix\doi{10.1186/s40679-019-0063-3}.
\bibitem[{Midgley and Eggeman(2015)}]{Midgley2015}
\bibinfo{author}{Midgley, P.A.}, \bibinfo{author}{Eggeman, A.S.},
  \bibinfo{year}{2015}.
\newblock \bibinfo{title}{Precession electron diffraction - a topical review}.
\newblock \bibinfo{journal}{IUCrJ} \bibinfo{volume}{2}, \bibinfo{pages}{126}.
\newblock \DOIprefix\doi{10.1107/S2052252514022283}.
\bibitem[{Nicoletti et~al.(2013)Nicoletti, de~la Pe{\~{n}}a, Leary, Holland,
  Ducati and Midgley}]{Nicoletti2013}
\bibinfo{author}{Nicoletti, O.}, \bibinfo{author}{de~la Pe{\~{n}}a, F.},
  \bibinfo{author}{Leary, R.K.}, \bibinfo{author}{Holland, D.J.},
  \bibinfo{author}{Ducati, C.}, \bibinfo{author}{Midgley, P.A.},
  \bibinfo{year}{2013}.
\newblock \bibinfo{title}{Three-dimensional imaging of localized surface
  plasmon resonances of metal nanoparticles}.
\newblock \bibinfo{journal}{Nature} \bibinfo{volume}{80}, \bibinfo{pages}{502}.
\newblock \DOIprefix\doi{10.1038/nature12469}.
\bibitem[{Nord et~al.(2020)Nord, Webster, Paton, McVitie, McGrouther, MacLaren
  and Paterson}]{Nord2020}
\bibinfo{author}{Nord, M.}, \bibinfo{author}{Webster, R.W.H.},
  \bibinfo{author}{Paton, K.A.}, \bibinfo{author}{McVitie, S.},
  \bibinfo{author}{McGrouther, D.}, \bibinfo{author}{MacLaren, I.},
  \bibinfo{author}{Paterson, G.W.}, \bibinfo{year}{2020}.
\newblock \bibinfo{title}{Fast pixelated detectors in scanning transmission
  electron microscopy. part i: Data acquisition, live processing, and storage}.
\newblock \bibinfo{journal}{Microsc. Microanal.} \bibinfo{volume}{26},
  \bibinfo{pages}{653--666}.
\newblock \DOIprefix\doi{10.1017/S1431927620001713}.
\bibitem[{Ophus(2019)}]{Ophus2019}
\bibinfo{author}{Ophus, C.}, \bibinfo{year}{2019}.
\newblock \bibinfo{title}{Four-dimensional scanning transmission electron
  microscopy (4{D}-{STEM}): from scanning nanodiffraction to ptychography and
  beyond}.
\newblock \bibinfo{journal}{Microscopy and Microanalysis} \bibinfo{volume}{25},
  \bibinfo{pages}{563}.
\newblock \DOIprefix\doi{10.1017/S1431927619000497}.
\bibitem[{Panova et~al.(2016)Panova, Chen, Bustillo, Ophus, Bhatt, Balsara and
  Minor}]{Panova2016}
\bibinfo{author}{Panova, O.}, \bibinfo{author}{Chen, X.C.},
  \bibinfo{author}{Bustillo, K.C.}, \bibinfo{author}{Ophus, C.},
  \bibinfo{author}{Bhatt, M.P.}, \bibinfo{author}{Balsara, N.},
  \bibinfo{author}{Minor, A.M.}, \bibinfo{year}{2016}.
\newblock \bibinfo{title}{Orientation mapping of semicrystalline polymers using
  scanning nanobeam diffraction}.
\newblock \bibinfo{journal}{Micron} \bibinfo{volume}{88}, \bibinfo{pages}{30}.
\newblock \DOIprefix\doi{10.1016/j.micron.2016.05.008}.
\bibitem[{Parish and Brewer(2010)}]{Parish2010}
\bibinfo{author}{Parish, C.M.}, \bibinfo{author}{Brewer, L.N.},
  \bibinfo{year}{2010}.
\newblock \bibinfo{title}{Multivariate statistics applications in phase
  analysis of {STEM}-{EDS} spectrum images}.
\newblock \bibinfo{journal}{Ultramicroscopy} \bibinfo{volume}{110},
  \bibinfo{pages}{134}.
\newblock \DOIprefix\doi{10.1016/j.ultramic.2009.10.011}.
\bibitem[{Pekin et~al.(2017)Pekin, Gammer, Ciston, Minor and Ophus}]{Pekin2017}
\bibinfo{author}{Pekin, T.C.}, \bibinfo{author}{Gammer, C.},
  \bibinfo{author}{Ciston, J.}, \bibinfo{author}{Minor, A.M.},
  \bibinfo{author}{Ophus, C.}, \bibinfo{year}{2017}.
\newblock \bibinfo{title}{Optimizing disk registration algorithms for nanobeam
  electron diffraction strain mapping}.
\newblock \bibinfo{journal}{Ultramicroscopy} \bibinfo{volume}{176},
  \bibinfo{pages}{170}.
\newblock \DOIprefix\doi{10.1016/j.ultramic.2016.12.021}.
\bibitem[{Pekin et~al.(2018)Pekin, Gammer, Ciston, Ophus and Minor}]{Pekin2018}
\bibinfo{author}{Pekin, T.C.}, \bibinfo{author}{Gammer, C.},
  \bibinfo{author}{Ciston, J.}, \bibinfo{author}{Ophus, C.},
  \bibinfo{author}{Minor, A.M.}, \bibinfo{year}{2018}.
\newblock \bibinfo{title}{In situ nanobeam electron diffraction strain mapping
  of planar slip in stainless steel}.
\newblock \bibinfo{journal}{Scripta Materalia} \bibinfo{volume}{146},
  \bibinfo{pages}{87}.
\newblock \DOIprefix\doi{10.1016/j.scriptamat.2017.11.005}.
\bibitem[{Ren et~al.(2018)Ren, Pueyo, Zhu, Debes and D{\^u}chene}]{Ren2018}
\bibinfo{author}{Ren, B.}, \bibinfo{author}{Pueyo, L.}, \bibinfo{author}{Zhu,
  G.B.}, \bibinfo{author}{Debes, J.}, \bibinfo{author}{D{\^u}chene, G.},
  \bibinfo{year}{2018}.
\newblock \bibinfo{title}{Non-negative matrix factorization: Robust extraction
  of extended structures}.
\newblock \bibinfo{journal}{The Astrophysical Journal} \bibinfo{volume}{852},
  \bibinfo{pages}{104}.
\newblock \DOIprefix\doi{10.3847/1538-4357/aaa1f2}.
\bibitem[{Ringe et~al.(2015)Ringe, DeSantis, Collins, Duchamp, Dunin-Barowski,
  Skrabalak and Midgley}]{Ringe2017}
\bibinfo{author}{Ringe, E.}, \bibinfo{author}{DeSantis, C.J.},
  \bibinfo{author}{Collins, S.M.}, \bibinfo{author}{Duchamp, M.},
  \bibinfo{author}{Dunin-Barowski, R.E.}, \bibinfo{author}{Skrabalak, S.E.},
  \bibinfo{author}{Midgley, P.}, \bibinfo{year}{2015}.
\newblock \bibinfo{title}{Resonances of nanoparticles with poor plasmonic metal
  tips}.
\newblock \bibinfo{journal}{Scientific Reports} \bibinfo{volume}{5},
  \bibinfo{pages}{17431}.
\newblock \DOIprefix\doi{10.1038/srep17431}.
\bibitem[{Rouviere et~al.(2013)Rouviere, B{\'e}ch{\'e}, Martin, Denneulin and
  Cooper}]{Rouviere2013}
\bibinfo{author}{Rouviere, J.L.}, \bibinfo{author}{B{\'e}ch{\'e}, A.},
  \bibinfo{author}{Martin, Y.}, \bibinfo{author}{Denneulin, T.},
  \bibinfo{author}{Cooper, D.}, \bibinfo{year}{2013}.
\newblock \bibinfo{title}{Improved strain precision with high spatial
  resolution using nanobeam precession electron diffraction}.
\newblock \bibinfo{journal}{Appl. Phys. Lett.} \bibinfo{volume}{103},
  \bibinfo{pages}{241913}.
\newblock \DOIprefix\doi{10.1063/1.4829154}.
\bibitem[{Savitzky et~al.(2020)Savitzky, Hughes, Zeltmann, Brown, Zhao, Pelz,
  Barnard, Donohue, DaCosta, Pekin, Kennedy, Janish, Schneider, Herring, Gopal,
  Anapolsky, Ercius, Scott, Ciston, Minor and Ophus}]{Savitzky2020}
\bibinfo{author}{Savitzky, B.H.}, \bibinfo{author}{Hughes, L.A.},
  \bibinfo{author}{Zeltmann, S.E.}, \bibinfo{author}{Brown, H.G.},
  \bibinfo{author}{Zhao, S.}, \bibinfo{author}{Pelz, P.M.},
  \bibinfo{author}{Barnard, E.S.}, \bibinfo{author}{Donohue, J.},
  \bibinfo{author}{DaCosta, L.R.}, \bibinfo{author}{Pekin, T.C.},
  \bibinfo{author}{Kennedy, E.}, \bibinfo{author}{Janish, M.T.},
  \bibinfo{author}{Schneider, M.M.}, \bibinfo{author}{Herring, P.},
  \bibinfo{author}{Gopal, C.}, \bibinfo{author}{Anapolsky, A.},
  \bibinfo{author}{Ercius, P.}, \bibinfo{author}{Scott, M.},
  \bibinfo{author}{Ciston, J.}, \bibinfo{author}{Minor, A.M.},
  \bibinfo{author}{Ophus, C.}, \bibinfo{year}{2020}.
\newblock \bibinfo{title}{{py4DSTEM}: a software package for multimodal
  analysis of four-dimensional scanning transmission electron microscopy
  datasets}.
\newblock \bibinfo{journal}{arXiv} \href{http://arxiv.org/abs/2003.09523}{{\tt
  arXiv:2003.09523}}.
\bibitem[{Savitzky et~al.(2019)Savitzky, Zeltmann, Barnard, DaCosta, Brown,
  Henderson, Ginsburg and Ophus}]{Savitzky2019}
\bibinfo{author}{Savitzky, B.H.}, \bibinfo{author}{Zeltmann, S.E.},
  \bibinfo{author}{Barnard, E.}, \bibinfo{author}{DaCosta, L.R.},
  \bibinfo{author}{Brown, H.G.}, \bibinfo{author}{Henderson, M.},
  \bibinfo{author}{Ginsburg, D.}, \bibinfo{author}{Ophus, C.},
  \bibinfo{year}{2019}.
\newblock \bibinfo{title}{py4dstem}.
\newblock \DOIprefix\doi{10.5281/zenodo.3333960}.
\bibitem[{Sunde et~al.(2018)Sunde, Marioara, van Helvoort and
  Holmestad}]{Sunde2017}
\bibinfo{author}{Sunde, J.K.}, \bibinfo{author}{Marioara, C.D.},
  \bibinfo{author}{van Helvoort, A.T.J.}, \bibinfo{author}{Holmestad, R.},
  \bibinfo{year}{2018}.
\newblock \bibinfo{title}{The evolution of precipitate crystal structures in an
  {A}l-{M}g-{S}i(-{C}u) alloy studied by a combined {HAADF}-{STEM} and {SPED}
  approach}.
\newblock \bibinfo{journal}{Materials Characterization} \bibinfo{volume}{142},
  \bibinfo{pages}{458}.
\newblock \DOIprefix\doi{10.1016/j.matchar.2018.05.031}.
\bibitem[{Uesugi et~al.(2020)Uesugi, Koshiya, Kikkawa, Nagai, Mitsuishi and
  Kimoto}]{Uesugi2020}
\bibinfo{author}{Uesugi, F.}, \bibinfo{author}{Koshiya, S.},
  \bibinfo{author}{Kikkawa, J.}, \bibinfo{author}{Nagai, T.},
  \bibinfo{author}{Mitsuishi, K.}, \bibinfo{author}{Kimoto, K.},
  \bibinfo{year}{2020}.
\newblock \bibinfo{title}{Non-negative matrix factorization for mining big data
  obtained using four-dimensional scanning transmission electron microscopy}.
\newblock \bibinfo{journal}{Ultramicroscopy} \bibinfo{volume}{221},
  \bibinfo{pages}{113168}.
\newblock \DOIprefix\doi{10.1016/j.ultramic.2020.113168}.
\bibitem[{Yaguchi et~al.(2008)Yaguchi, Konno, Kamino and
  Watanabe}]{Yaguchi2008}
\bibinfo{author}{Yaguchi, T.}, \bibinfo{author}{Konno, M.},
  \bibinfo{author}{Kamino, T.}, \bibinfo{author}{Watanabe, M.},
  \bibinfo{year}{2008}.
\newblock \bibinfo{title}{Observation of three-dimensional elemental
  distributions of a {S}i device using a 360$^\circ$-tilt {FIB} and the cold
  field-emission {STEM} system}.
\newblock \bibinfo{journal}{Ultramicroscopy} \bibinfo{volume}{108},
  \bibinfo{pages}{1603}.
\newblock \DOIprefix\doi{10.1016/j.ultramic.2008.06.003}.
\bibitem[{Zeltmann et~al.(2020)Zeltmann, M{\"u}ller, Bustillo, Savitzky,
  Hughes, Minor and Ophus}]{Zeltmann2020}
\bibinfo{author}{Zeltmann, S.E.}, \bibinfo{author}{M{\"u}ller, A.},
  \bibinfo{author}{Bustillo, K.C.}, \bibinfo{author}{Savitzky, B.},
  \bibinfo{author}{Hughes, L.}, \bibinfo{author}{Minor, A.M.},
  \bibinfo{author}{Ophus, C.}, \bibinfo{year}{2020}.
\newblock \bibinfo{title}{Patterned probes for high precision {4D-STEM} {B}ragg
  measurements}.
\newblock \bibinfo{journal}{Ultramicroscopy} \bibinfo{volume}{209},
  \bibinfo{pages}{112890}.
\newblock \DOIprefix\doi{10.1016/j.ultramic.2019.112890}.

\end{thebibliography}

\end{document}


\begin{frontmatter}

\title{Supplementary Materials for:\\Fast Grain Mapping with Sub-Nanometer Resolution Using 4D-STEM with Grain Classification by Principal Component Analysis and Non-Negative Matrix Factorization}

\author[label1,label2]{Frances I. Allen\corref{cor1}}

\cortext[cor1]{Corresponding author}
\ead{francesallen@berkeley.edu} 

\address[label1]
{Department of Materials Science and Engineering, UC Berkeley, Berkeley, CA 94720, USA}
\address[label2]
{National Center for Electron Microscopy, Molecular Foundry, LBNL, Berkeley, CA 94720, USA}

\author[label1,label2]{Thomas C. Pekin\fnref{fn1}}
\fntext[fn1]{Current address: Institut für Physik \& IRIS Adlershof, Humboldt Universität zu Berlin, 12489 Berlin, Germany}

\author[label3]{Arun Persaud}
\address[label3]{Accelerator Technology and Applied Physics Division, LBNL, Berkeley, CA 94720, USA}

\author[label4]{Steven J. Rozeveld}
\author[label4]{\\Gregory F. Meyers}
\address[label4]
{Core R\&D - Analytical Sciences, The Dow Chemical Company, Midland, MI 48674, USA\vskip 10pt \textnormal{\today}}

\author[label2]{Jim Ciston}
\author[label2]{Colin Ophus}
\author[label1,label2]{Andrew M. Minor}

\end{frontmatter}

\newpage

\begin{figure*}
\centering
\includegraphics[width=0.9\linewidth]{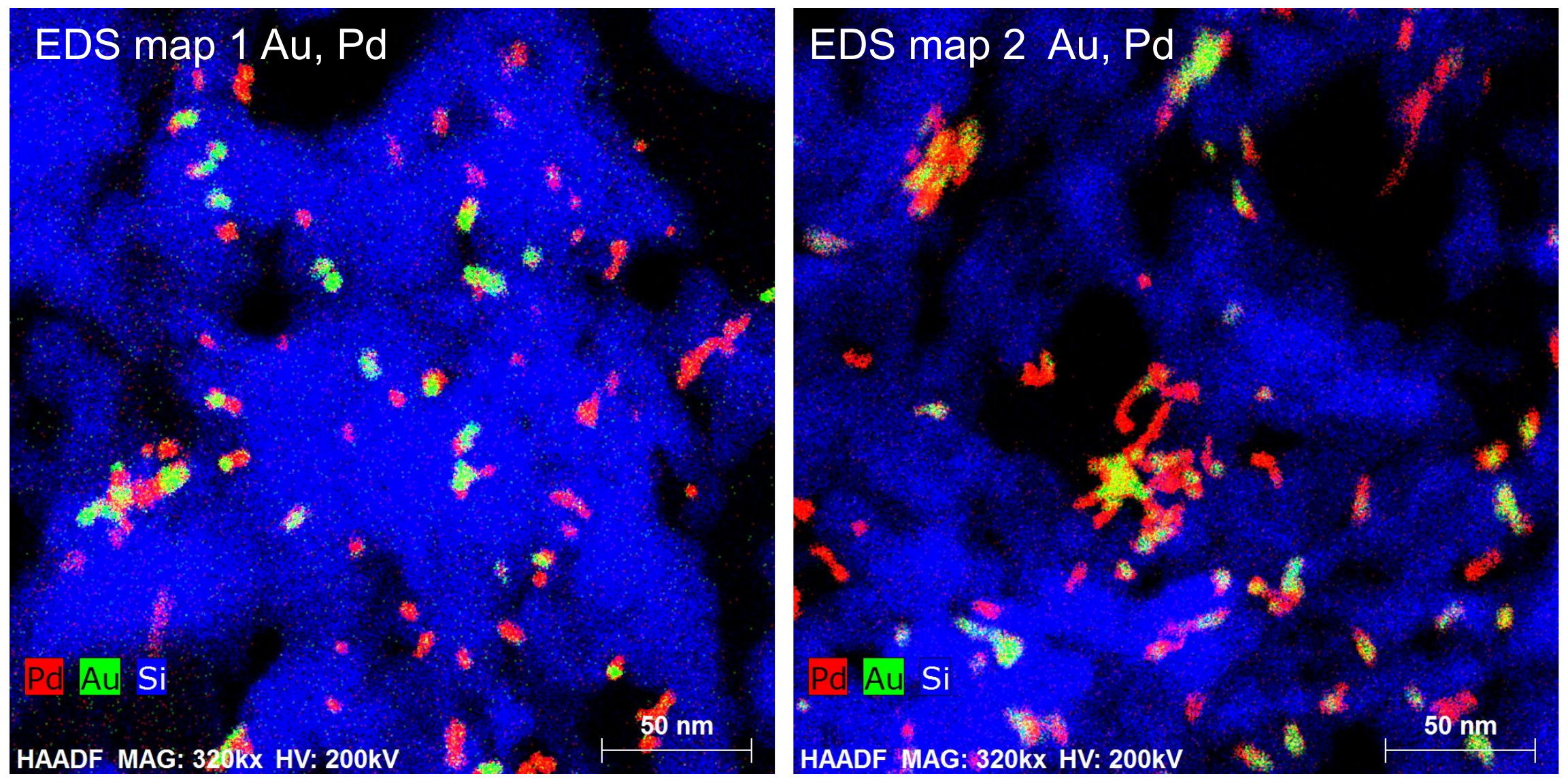}
  \caption{STEM-XEDS elemental maps for gold, palladium and silicon for the gold-palladium nanoparticles on the silica support.}
  \label{FigS1}
\end{figure*}

\begin{figure*}
\centering
\includegraphics[width=0.9\linewidth]{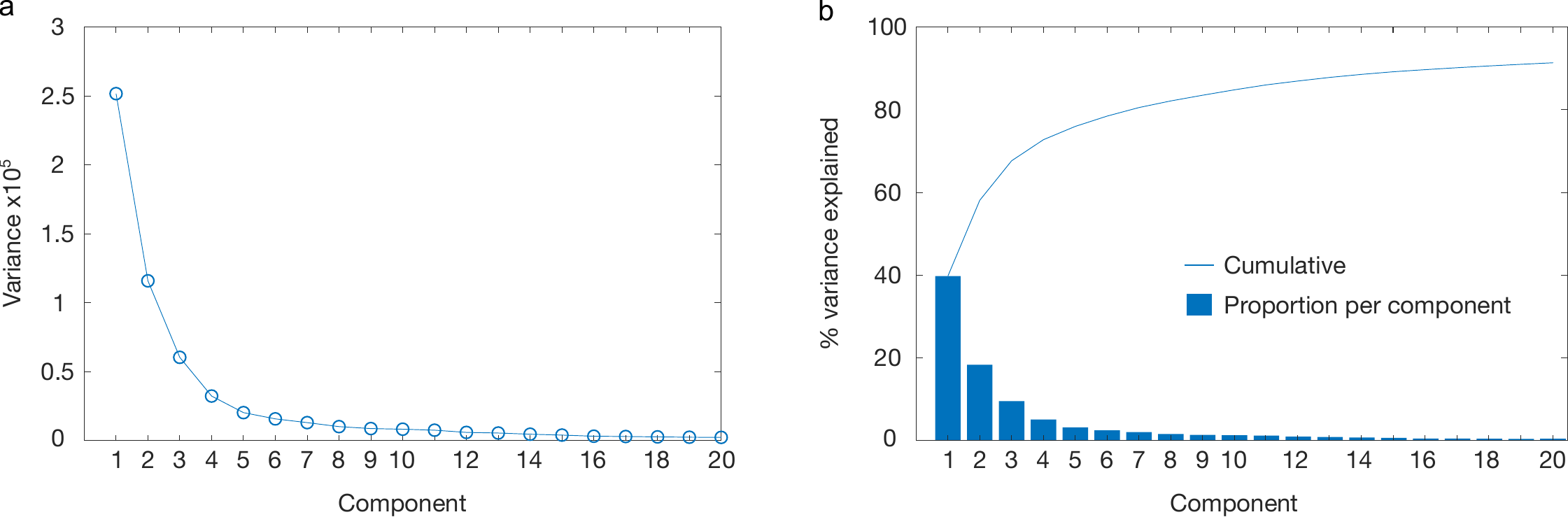}
  \caption{PCA grain mapping results for a single nanoparticle cluster showing (a) the scree plot and (b) a plot of the percentage of the total variance explained for the first 20 components.}
  \label{FigS2}
\end{figure*}

\begin{figure*}
\centering
\includegraphics[width=0.9\linewidth]{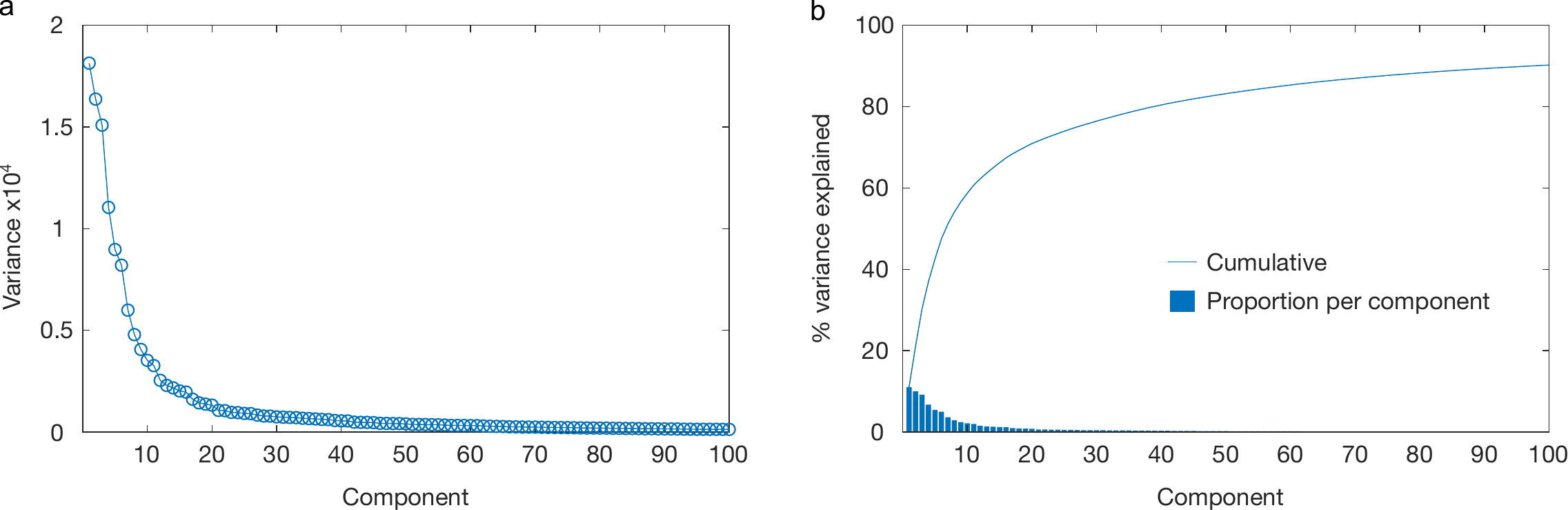}
  \caption{PCA grain mapping results for a distribution of nanoparticles showing (a) the scree plot and (b) a plot of the percentage of the total variance explained for the first 100 components.}
  \label{FigS3}
\end{figure*}
